\documentclass[runningheads]{llncs}
\usepackage[T1]{fontenc}
\usepackage{wrapfig}
\usepackage{graphicx}
\usepackage{amsmath}
\usepackage{amssymb}
\usepackage{tabularx}
\usepackage{booktabs}
\usepackage{adjustbox}
\usepackage{xcolor}
\usepackage{bm}

\usepackage{mathtools}
\usepackage{esvect}
\usepackage{longtable}
\usepackage{multirow}
\usepackage{array}
\usepackage{rotating}
\usepackage{subcaption} 
\usepackage{threeparttable}
\usepackage{calc}
\usepackage{algorithm}
\usepackage{algorithmic}
\usepackage{setspace}  
\usepackage{bbm}
\usepackage{colortbl}
\usepackage[hidelinks]{hyperref}
\usepackage{color}

\usepackage{cleveref}

\newcommand{\SetOfLocationsIncludePickItem}[1]{\mathcal{V}_{\!#1}^\text{S}}
\newcommand{\SetOfSKUs}{\mathcal{P}}
\newcommand{\SetOfAllNodes}
{\mathcal{V}}
\newcommand{\SetOfPackingStations}{\mathcal{V}^{\text{D}}}
\newcommand{\SetOfStorageLocations}{\mathcal{V}^{\text{S}}}
\newcommand{\NumTours}{\mathcal{B}}
\newcommand{\SetOfShelves}{\mathcal{V}^{\text{R}}}
\newcommand{\SetOfAgents}{\mathcal{M}}

\newcommand{\logits}{L}
\newcommand{\curragent}[1]{{\Omega_{#1}}}
\newcommand{\PermOverL}{{\Omega(\logits)}}

\newcommand{\shelfPolicy}{g^{\mathcal{V}}_{\theta}}
\newcommand{\SKUPolicy}{g^{\mathcal{P}}_{\theta}}

\newcommand{\numagents}{{M}}
\newcommand{\agentidx}{{m}}

\newcommand{\embdim}{{D}}

\begin{document}
\vspace{-2cm}
\title{Learning to Solve the Min-Max \\ Mixed-Shelves Picker-Routing Problem \\ via Hierarchical and Parallel Decoding}
\authorrunning{Luttmann, Xie}
\titlerunning{Hierarchical and Parallel Decoding for Picker-Routing}
%

%
\author{Laurin Luttmann\inst{1} \and
Lin Xie\inst{2}}
%

%
\institute{Leuphana University, Lüneburg, Germany \and
Brandenburg University of Technology, Cottbus, Germany}

\maketitle              
\begin{abstract}
The Mixed-Shelves Picker Routing Problem (MSPRP) is a fundamental challenge in warehouse logistics, where pickers must navigate a mixed-shelves environment to retrieve SKUs efficiently. Traditional heuristics and optimization-based approaches struggle with scalability, while recent machine learning methods often rely on sequential decision-making, leading to high solution latency and suboptimal agent coordination. In this work, we propose a novel hierarchical and parallel decoding approach for solving the min-max variant of the MSPRP via multi-agent reinforcement learning. While our approach generates a joint distribution over agent actions, allowing for fast decoding and effective picker coordination, our method introduces a sequential action selection to avoid conflicts in the multi-dimensional action space. Experiments show state-of-the-art performance in both solution quality and inference speed, particularly for large-scale and out-of-distribution instances. Our code is publicly available at \url{http://github.com/LTluttmann/marl4msprp}

\end{abstract}
\section{Introduction}
Order picking, the process of retrieving items from a warehouse to fulfill customer orders, is one of the most labor-intensive and time-consuming operations in warehouse logistics, accounting for up to 65\% of total operating costs \cite{de2007design}. In conventional picker-to-parts warehouses, most of a picker's time is spent traveling between the shelves of the storage area \cite{Tompkins2010}. To reduce travel time, mixed-shelves storage strategies have gained traction in recent years (see \cite{Boysen2017}, \cite{weidingerPickerRoutingMixedshelves2019}, \cite{xie2021introducing}, \cite{xie2023formulating}, and \cite{luttmann2024neural}). Unlike traditional warehouse layouts that allocate a single storage position per Stock-Keeping Unit (SKU), mixed-shelves storage distributes SKUs to multiple shelves of the storage area, potentially decreasing travel distances and improving overall efficiency.

This mixed-shelves approach gives rise to the Mixed-Shelves Picker Routing Problem (MSPRP), which focuses on determining optimal routes for pickers while considering the unique constraints of mixed-shelves warehouses and operations. Despite its practical significance, research on solving the MSPRP remains limited. Existing approaches primarily rely on classical heuristics such as variable neighborhood search \cite{xie2023formulating} and Tabu search \cite{danielsModelWarehouseOrder1998}. While these methods can produce high-quality solutions, they are computationally expensive which makes them impractical for large-scale or real-time applications.
Neural combinatorial optimization (NCO) has emerged as a promising alternative, offering faster solution generation while maintaining high solution quality across various routing problems. However, current NCO applications to the MSPRP are limited to single-picker scenarios that focus on minimizing total travel distance. This represents a significant gap in addressing real-world warehouse operations, where multiple pickers typically work simultaneously and minimizing the longest tour (i.e., the route of the most time-consuming picker) is more critical for maintaining efficient operations. To bridge this gap, we propose a novel NCO approach that integrates hierarchical and parallel decoding to efficiently solve the min-max variant of the MSPRP. Our main contributions are as follows:

\begin{itemize}
    \item We formulate the MSPRP as a cooperative multi-agent problem, aiming to balance workloads among pickers rather than minimizing the total distance.
    \item A Hierarchical and Parallel Decoding framework enables efficient picker coordination in complex multi-dimensional action spaces.
    \item A Sequential Action Selection strategy supports the parallel decoding step by avoiding conflicts while exhibiting strong generalization performance    
    \item We demonstrate state-of-the-art performance in terms of both solution quality and computational efficiency, particularly for large problem instances.
\end{itemize}

\section{Related Work}
\label{subsec:relwork}

\subsubsection{Mixed-shelves Picker Routing.}
Various heuristics have been developed to address the MSPRP, including construction and improvement methods \cite{weidingerPickerRoutingRectangular2018,weidingerPickerRoutingMixedshelves2019} and a variable neighborhood search approach \cite{xie2023formulating}. However, these methods often require minutes of computation, which can be impractical for fast-paced operations.
Only one neural learning approach has been proposed for the MSPRP, modeling it as a heterogeneous graph to optimize selection and routing for a single picker \cite{luttmann2024neural}. In practice, multiple pickers operate simultaneously, shifting the focus towards minimizing overall completion time instead of total travel distance. We thus explore a min-max variant of the MSPRP, aiming to balance travel distances among pickers. 
\vspace{-5mm}

\subsubsection{Neural Combinatorial Optimization.} While early work in the NCO field focus on problems involving a single agent like in the traveling salesman problem  \cite{vinyals2015pointer,kool2018attention,kwon2020pomo,kwon2021matrix}, recently more attention has been given to more complex, multi-agent variants of routing problems. Building on \cite{kool2018attention}, the Equity Transformer \cite{son2024equity} and 2d-Ptr \cite{liu20242d} introduce attention-based policies for multi-agent min-max routing. However, these models can be seen as purely autoregressive approaches, constructing solutions for one agent at a time, thus neglecting potential agent coordination and exhibiting high generation latency for large problems with many agents. PARCO \cite{berto2024parco} aims to address these shortcomings by introducing parallel solution construction, using a Priority-based Conflict handler to avoid infeasible solutions when performing actions for multiple agents simultaneously. 

In this work, we combine the hierarchical decoder of \cite{luttmann2024neural}, designed for the integrated selection and routing in MSPRP, as well as parallel solution construction similar to \cite{berto2024parco}, to learn high quality solutions for the min-max MSPRP. To effectively avoid conflicts during hierarchical solution construction, we combine a Parallel Pointer Mechanism with a Sequential Action Selection algorithm.

\section{Problem Formulation}
\label{sec:mathmod}

This work focuses on a min-max variant of the MSPRP with split orders and split deliveries covered in \cite{luttmann2024neural}. The split orders assumption allows items of an order to be picked
within different tours and split deliveries relaxes the assumption that the demand
for an SKU must be satisfied by a single picker tour \cite{xie2021introducing}. A tour is defined by the storage locations visited between two successive visits to a packing station $h \in \mathcal{V}^{\mathrm{D}}$, where picked items are unloaded and commissioned. During a tour, no more than $\kappa$ units can be picked. Further, due to the mixed-shelves storage policy each shelf may consist of multiple storage locations or compartments storing units of different SKUs. Also, the mixed-shelves storage policy allows each SKU $p$ to be retrieved from multiple storage locations $i \in \SetOfLocationsIncludePickItem{p}$. 

The goal of the min-max MSPRP is to pick all $d_p$ demanded units of all requested SKUs $p \in \SetOfSKUs$ and returning them to a packing station $h \in \SetOfPackingStations$ while minimizing the maximum travel distance among the individual pickers $m = 1,...,M$, henceforth also called agents. 
Note that in order to compare our proposed method against baselines \cite{luttmann2024neural}, \cite{liu20242d}, and \cite{son2024equity}, we assume that the number of agents is equal to the number of tours required to collect all demanded items given the picker capacity $\kappa$ (i.e. $\numagents = \left \lceil \frac{\sum_p d_p}{\kappa} \right \rceil$). We provide the mathematical model for the min-max MSPRP in \Cref{appendix:notation}.

\subsection{Markov Decision Process Formulation}
\label{sec:mdp}
\newcommand{\setOfNodesMDP}{\mathcal{V}}
The min-max MSPRP can be modeled as a cooperative Multi-Agent Markov Decision Process (MMDP) with $\numagents$ agents sharing a common reward. An MMDP is defined as $(\mathcal{S}, \SetOfAgents, \{\mathcal{A}_i\}, \Gamma, R)$, where $\mathcal{S}$ and $\SetOfAgents$ are finite sets of states and agents, respectively. Each agent $m$ selects actions from $\mathcal{A}_m$, with the joint action space denoted as $\bm{\mathcal{A}}$. The transition function $\Gamma$ determines state changes based on actions, and $R$ is the shared reward function.  

MMDPs involve sequential decision-making, where agents select and execute actions at each step until a terminal state $s_T$ is reached. The min-max MSPRP is framed as an MMDP, with pickers (agents) visiting warehouse shelves to fulfill SKU demands. A shared $\theta$-parameterized policy determines the next location and SKU to pick. This chapter formally defines the min-max MSPRP as an MMDP, specifying its state, action space, transition rule, and reward function.

\vspace{-4.5mm}
\subsubsection{State.} 
The state $s_t$ of the min-max MSPRP at step $t$ can be represented as a heterogeneous graph $\mathcal{G}=(\setOfNodesMDP,\SetOfSKUs, \SetOfAgents, E_{t})$ with pickers, warehouse locations and SKUs posing different types of nodes in the graph. 
The set of warehouse locations $\setOfNodesMDP$ is the collection of all packing stations $\SetOfPackingStations$ and shelves $\mathcal{V}^\text{R}$. 
The state of an SKU $p \in \SetOfSKUs$ is defined by its remaining demand $d_{pt}$ at step $t$. 
Moreover, edges with weights $E_t$ connect shelf and SKU nodes, specifying the storage quantity $e_{vpt}$ of an item $p$ in the respective shelf $v$ at time $t$. 
Lastly, the state of the pickers $m \in \SetOfAgents$ is defined by their current location $v^m_{t}$, remaining capacity $\kappa^m_{t}$ and the length of their current tour $\tau^m_{1:t}=(v_1^m,\ldots,v_t^m)$, denoted as $dist(\tau^m_{1:t})$.
\vspace{-4.5mm}
\subsubsection{Action.} 
A single agent action $a^m_{t}$ is a tuple $(v,p)$ specifying the next shelf to visit as well as the SKU to pick for agent $m$. Given $s_t$, visiting shelf $v$ is a feasible action if it stores items of at least one SKU currently in demand. Furthermore, given the picking location $v$, the picker may only select an SKU for picking that is both still in demand and available in the current shelf. Note, that the quantity of picked items will be determined heuristically by the transition function $\Gamma$ in order to decrease the complexity of the action space and facilitate policy learning. 

The packing station can always be visited by a picker to unload picked items and thus to restore the capacity. When a picker's capacity is exhausted, visiting the packing station is the only possible action. Moreover, to facilitate agent coordination, a picker may always choose to stay at its current location in order to give other pickers precedence. This way, a hesitant picker may wait and evaluate what the other pickers are doing, before making the next move. 
\vspace{-4.5mm}
\subsubsection{Transition.} 
Given the joint actions $\bm{a}_t = (a_t^1,\ldots,a_t^m)$ of all agents, the transition function $\Gamma(s_t, \bm{a}_t)$  deterministically transits to $s_{t+1}$. The new state consists of the updated agent locations $\bm{v}_{t} = (v^1_{t}, \ldots, v^\numagents_{t})$ and agent tours $\tau^m_{1:t} = \tau^m_{1:t-1} \cup{\{ v^m_{t} \}}$.
To update the remaining demand, supply and picker capacities, the pick quantity $y_t^m$ must be determined. Given the pick locations, SKUs and a permutation $\Omega$ over pickers, we iteratively determine the pick quantity as the minimum of the remaining demand of the selected SKU $p$, the storage quantity at the agent's new location $v$ as well as the agent's remaining capacity:
\begin{align}
    y^\curragent{k}_t = \text{min} (\kappa_t^\curragent{k}, \, d_{pt} - \sum_{j=1}^{k-1} y_t^\curragent{j} \cdot \mathbbm{1}_{p^{\curragent{j}}_t = p}, \, e_{vpt} ),
\end{align}
where no two agents may select the same shelf-SKU combination in the same decoding step, for which reason the supply $e_{vpt}$ will not be altered by preceding agents. 
Given the pick quantities $y_t^m$, the transition function updates the demand $d_{pt+1} = d_{pt} - \sum_{m=1}^\numagents y^m_{t} \cdot \mathbbm{1}_{p^m_t = p}$, the supply $e_{vpt+1} = e_{vpt} - \sum_{m=1}^\numagents y^m_{t} \cdot \mathbbm{1}_{v^m_{t} = v; p^m_t = p}$ and the remaining picker capacity $\kappa^m_{t+1} = \kappa^m_{t} - y^m_{t}$. 
The problem instance $x$ is solved once the demand for every SKU is met and all pickers have returned to the packing station they were starting from. A feasible solution to $x$, reaching the terminal state $s_T$ in $T$ construction steps, will be denoted as $\bm{a} \coloneqq (\bm{a}_1, \ldots, \bm{a}_{T})$. 
\vspace{-4.5mm}
\subsubsection{Reward.}
The MMDP formulation of the min-max MSPRP has a sparse reward function, which is only defined for a complete solution $\bm{a}$. We define the reward $R(\bm{a}, x)$ as the negative of the maximum travel distance of any picker, i.e. $R(\bm{a},x) = - \text{max}_{m \in \SetOfAgents} \, dist(\tau^m_{1:T})$, and the goal of our approach is to maximize it.

\section{Method}


This section introduces our Multi-Agent Hierarchical Attention Model (MAHAM) -- an extension of the Hierarchical Attention Model (HAM) architecture \cite{luttmann2024neural} -- designed to address the multi-picker min-max variant of the MSPRP. In NCO, the sequential nature of the MDP underlying the CO problem often leads to the adoption of autoregressive (AR) models, which implement a sequential solution generation via an encoder-decoder network, formally represented as:\footnote{henceforth, we use the current problem state $s_t$ instead of the problem instance $x$ and the previous actions $\bm{a}_{1:t-1}$ to condition the models}
\begin{align}
\label{eq:autoregressive_encoding_decoding}
p_\theta(\bm{a}|x) &= \prod_{t=1}^{T} g_\theta(a_{t} | x, \bm{a}_{1:t-1}, H_t) \cdot f_\theta(H_t | x, \bm{a}_{1:t-1})
\end{align}
where $f_\theta$ represents the encoder network, used to construct a hidden representation of the problem instance $x$ given the actions taken so far and $g_\theta$ the decoder, that selects actions based on the problem encoding and its current state. 

\begin{figure*}[t]
    \centering
    \includegraphics[width=\textwidth]{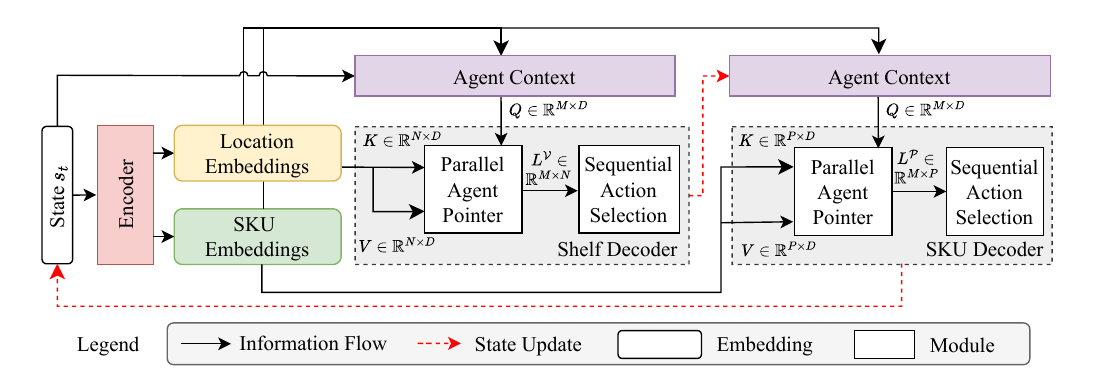}
    \vspace{-5mm}
    \caption{Overview of the MAHAM Architecture}
    \vspace{-5mm}
    \label{fig:maham}
\end{figure*}

MAHAM follows this approach, however the presence of multiple agents and a composite action space $\mathcal{A} \equiv \mathcal{V} \times \mathcal{P}$ introduce special needs which we carefully address with our architecture in \Cref{fig:maham}. While existing approaches tackle multi-agent problems by sequentially generating solutions for one agent after another \cite{son2024equity} or using a separate decoder $\pi_\theta^m$ per agent \cite{zong2022mapdp}, MAHAM poses a shared policy which constructs multiple picker routes in parallel through 1.) a separate agent encoder and 2.) a parallel decoding with sequential action selection scheme.  

\subsection{Encoder}

\subsubsection{Problem Encoder.}
\label{sec:encoder}

As defined earlier, the min-max MSPRP can be represented as a heterogeneous graph with agents, packing stations, shelves and SKUs posing different node types. We follow \cite{luttmann2024neural} and first project these different node-types from their distinct feature spaces into a mutual embedding space of dimensionality $\embdim$ using type-specific transformations $W_{\phi_i}$ for node $i$ of type $\phi_i$. The features used to represent agents, stations, shelves and SKUs in the features space are listed in \Cref{tab:features} in \Cref{appendix:network}.

Also similar to \cite{luttmann2024neural}, we use several layers of self- and cross-attention between location and SKU nodes. To this end, we treat packing stations as shelves that store zero units for each SKU and concatenate their initial embeddings to those of the shelf nodes, yielding $H^{0}_\mathcal{V} = [H^{0}_{\SetOfPackingStations} || H^{0}_{\SetOfShelves}]$. Likewise, the initial SKU embeddings are denoted as $H^{0}_\mathcal{P}$. While self-attention is applied independently to shelf and SKU embeddings following the Transformer architecture \cite{vaswaniAttentionAllYou2017a}, cross-attention allows shelves and SKUs to influence each other’s embeddings. Consequently, shelf embeddings encode information about the SKUs they store, and SKU embeddings reflect their placement within the storage area -- an essential property for hierarchical action selection. To perform cross-attention we compute a single matrix of attention scores $A$ using shelf embeddings as queries $Q$ and SKU embeddings as keys $K$. This contrasts with the MatNet \cite{kwonMatrixEncodingNetworks2021a} and HAM \cite{luttmann2024neural} architectures, which compute separate attention scores for each node type—once as queries and once as keys. Formally we perform:
\begin{align}
    \label{eq:matnet_dot_score}
    A &= \frac{QK^\top}{\sqrt{d_k}}, \qquad Q = W^Q H_{\mathcal{V}}^{l-1}, \qquad  K = W^K H_{\mathcal{P}}^{l-1}
\end{align}
where $W^Q$ and $W^K \in \mathbb{R}^{d_k \times \embdim}$ are weight matrices learned per attention head\footnote{For succinctness, we omit the layer and head enumeration} and $d_k$ is the per-head embedding dimension. The resulting attention scores $A\in \mathbb{R}^{|\mathcal{V}| \times |\mathcal{P}|}$ can be interpreted as the (learned) influence of an SKU $p$ on the embedding of location $v$. 
Similar to MatNet \cite{kwonMatrixEncodingNetworks2021a} we fuse these learned attention scores with the supply-matrix $E \in \mathbb{R}^{|\mathcal{V}| \times |\mathcal{P}|}$, which specifies how many units of SKU $p$ are stored in location $v$. To this end, we concatenate the attention score and the matrix of storage quantities and feed the resulting score vector through a multi-layer perceptron $\text{MLP}: \mathbb{R}^{|\mathcal{V}| \times |\mathcal{P}| \times 2} \rightarrow \mathbb{R}^{|\mathcal{V}| \times |\mathcal{P}|}$, with a single hidden layer comprising of $\embdim$ units and GELU activation function \cite{hendrycks2016gaussian}. Further, we pass the transpose of the attention scores and of the supply matrix $A^\top, \, E^\top \in \mathbb{R}^{|\mathcal{P}| \times |\mathcal{V}|}$ through a second MLP to obtain the influence $A_{\mathcal{P} \rightarrow \mathcal{V}}$ of locations $v$ on the SKU embeddings $H_{\mathcal{P}}$:
\begin{align}
    \label{eq:matnet_mixed_score}
     A_{\mathcal{V} \rightarrow \mathcal{P}} = \mathrm{MLP}_{\mathcal{V}} \big ([A || E ] \big ), \quad  A_{\mathcal{P} \rightarrow \mathcal{V}} = \mathrm{MLP}_{\mathcal{P}} \big ([A^\top || E^\top ] \big ),
\end{align}
By avoiding to compute the (computationally expensive) attention scores twice, once to generate shelf embeddings and once for the SKU embeddings, our implementation of the cross-attention mechanism leverages parameter sharing, improving both efficiency and generalization performance, as demonstrated in \Cref{sec:exp}. 
The resulting attention scores are then used to compute the embeddings for the nodes of the respective type: 
\begin{align}
     H_{\mathcal{V}}^\prime &= \text{softmax}(A_{\mathcal{V} \rightarrow \mathcal{P}})V_\mathcal{P}, \quad V_\mathcal{P} = W^{V}_\mathcal{P} H_{\mathcal{P}}^{l-1} \\
     H_{\mathcal{P}}^\prime &= \text{softmax}(A_{\mathcal{P} \rightarrow \mathcal{V}})V_\mathcal{V}, \quad V_\mathcal{V} = W^{V}_\mathcal{V} H_{\mathcal{V}}^{l-1}
\end{align}
As in \cite{vaswaniAttentionAllYou2017a}, $H_{\mathcal{V}}^\prime$ and $H_{\mathcal{P}}^\prime$  are then augmented through skip connections, layer normalization, and a feed-forward network, yielding the location and SKU embeddings $H_{\mathcal{V}}^l$ and $H_{\mathcal{P}}^l$, respectively, of the current layer $l$. 
%

\setlength{\intextsep}{10pt}%
\begin{wrapfigure}{r}{0.6\textwidth}
  \begin{center}
    \vspace{-20pt}
    \includegraphics[width=0.6\textwidth]{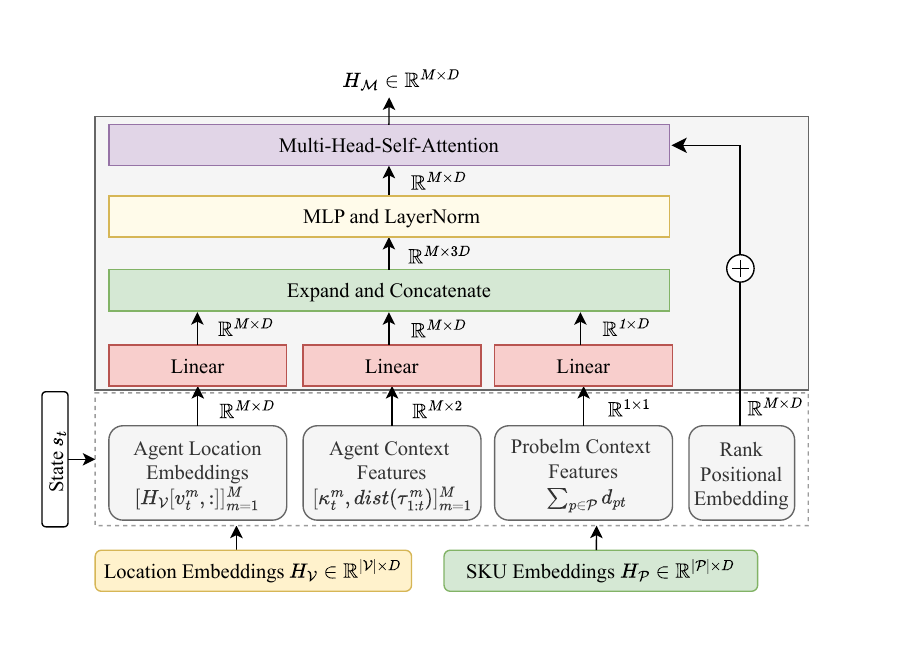}
    \vspace{-35pt}
  \end{center}
  \caption{Agent Context Encoder}
  \label{fig:cntxt}
\end{wrapfigure}

\subsubsection{Agent Encoder.}
To account for multiple agents, we introduce an Agent Context Encoder, as illustrated in \Cref{fig:cntxt}, into our MAHAM architecture. This encoder leverages the embeddings $H_\mathcal{V}$ and $H_\mathcal{P}$ from the problem encoder, along with the current state $s_t$, to generate embeddings for each picker.  
To facilitate informed decision-making at each decoding step, the agent embeddings incorporate three key types of information. First, spatial information of pickers is captured by using the embedding of a picker's current location. 
Further, the remaining capacity and the length of an agent's current tour are included in the agent encoder, helping the model to determine whether to continue the tour or send the picker to a packing station.
Lastly, the total demand across all SKUs and the average-pooled SKU embeddings provide insights into the remaining workload. 
Each context feature is first projected into a shared embedding space of dimensionality $\embdim$. The resulting embeddings are then concatenated and passed through an MLP, ensuring that the final representation is mapped back to the original embedding space.

Since coordination between pickers is critical in the min-max MSPRP, we add a Multi-Head-Self-Attention (MHSA) layer \cite{vaswaniAttentionAllYou2017a} at the end of the Agent Context Encoder, which enables message passing between agents. As in \cite{vaswaniAttentionAllYou2017a}, we add a positional encoding to the agent embeddings before they enter the MHSA layer. However, given the absence of a natural ordering of pickers, we employ a Ranking-based Position Encoding, where pickers are ranked in descending order of their remaining capacity. This allows the encoder to better prioritize agents based on their current workload, which is crucial for the sequential action selection that will be described in \Cref{sec:seq_action_selection}. We denote the final agent embeddings as $H_\SetOfAgents$ and the set of all embeddings as $H=(H_\mathcal{V},H_\mathcal{P},H_\SetOfAgents)$.

\subsection{Parallel and Hierarchical Decoder}

Given the embeddings for warehouse locations, SKUs, and agents from the encoder, the decoder determines the next location to visit by the pickers as well as the SKUs to be picked there. In contrast to other architectures like \cite{son2024equity}, our approach generates trajectories for all $\numagents$ simultaneously. This way, the agents can coordinate and balance the workload. If, for example, in step $t$ the tour of picker $m=1$ is much longer than that of picker $m=2$, the agents can coordinate that $m=2$ picks an SKU that is only available in far away shelves. This kind of coordination is not possible in purely autoregressive settings, where agent trajectories are constructed one after another. 

For our parallel and hierarchical decoding scheme, we adopt the hierarchical decoder architecture from \cite{luttmann2024neural} to sample actions specifying the next locations $v^{\agentidx}_{t}$ to visit and the SKU $p^{\agentidx}_{t}$  to pick by agents $\agentidx=1,\ldots \numagents$. To this end, we define two decoders $\shelfPolicy: \mathcal{S} \rightarrow \mathcal{V}$ and $\SKUPolicy: \mathcal{S} \rightarrow \mathcal{P}$ for action subspaces $\setOfNodesMDP$ and $\SetOfSKUs$, respectively. Moreover, we define a partial transition function $s_t^\prime = \Gamma_\circ(s_t, \bm{v}_t)$, generating an intermediate state $s_t^\prime$ with updated location information. The decoders can then be used in a hierarchical manner to generate the joint probability of a shelf-SKU pair according to the chain rule of probability \cite{luttmann2024neural}:
\begin{align}
    \label{eq:multi-agent-policy}
    g_\theta (\bm{a}_t \, | \, s_t, H) = \shelfPolicy (\bm{v}_t \, | \,  s_t, H) \cdot \SKUPolicy (\bm{p}_t \, | \, s_t^\prime, H),
\end{align}
where $\bm{v}_t$ and $\bm{p}_t$ are the joint agent actions for the shelf and SKU sub-action spaces, respectively. MAHAM models the joint agent actions $\bm{a}_{dt}$ for each sub-action space $d$ and the corresponding decoder $g_\theta^d$ as an autoregressive sequence generation process, similar to \Cref{eq:autoregressive_encoding_decoding}:
\begin{align}
\label{eq:maham}
p_\theta(\bm{a}_{dt} \, | s_t, \, H) &= g^d_\theta(\logits_d \, | \, s_t, H) \cdot \prod_{i=1}^{M} \psi \left( a^{\PermOverL_{i}}_{dt} | \logits_d, \bm{a}^{\PermOverL_{1:i-1}}_{dt} \right)
\end{align}
where $\logits_d \in \mathbb{R}^{\numagents \times |\mathcal{A}_d|}$ are the unnormalized log-probabilities (henceforth logits) over the joint action space generated by sub-policy $g_\theta^d$ and $\PermOverL$ is a permutation over agents given $\logits_d$. Further, $\psi$ is a stochastic function, which autoregressively selects actions based on the logits $L$ as well as the action sequence $\bm{a}^{\PermOverL_{1:i-1}}_{dt}$ of preceding agents.
Note, that while the policy $p_\theta$ itself acts autoregressively according to the sequential action selection strategy $\psi$, \Cref{eq:maham} factors out the computationally expensive calculation of the log-probabilities, which is done in parallel for all agents, allowing an effective and efficient agent coordination and ranking. Therefore, both the shelf and SKU decoder are modifications of the AM decoder proposed by \cite{kool2018attention}, which uses the cross-attention mechanism to generate unnormalized log-probabilities $\bm{l} \in \mathbb{R^{|\mathcal{A}|}}$. However, in contrast to \cite{kool2018attention}, who use a single context vector as query, our architecture uses the agent embeddings $H_{\SetOfAgents} \in \mathbb{R}^{\numagents \times D}$ in the cross-attention mechanism: 
\begin{align}
    \label{eq:attn_dec}
    Q_d &= \text{Attn}(H_{\SetOfAgents} W^Q, \, H_d W^K, \, H_d W^V) \\
    \logits_d &= C \cdot tanh \left ( \frac{Q_d K_d^\top}{\sqrt{\embdim}}  \right )
\end{align}
where $C$ is a scale parameter, $K_d$ is a projection of the embeddings $H_d$ belonging to the sub-action space $d$ of the decoder. In the following we will show how we can use the logits of the joint action space $\logits_d \in \mathbb{R}^{\numagents \times |\mathcal{A}_d|}$ to generate feasible actions $\bm{a}_{dt}=(a^{1}_{dt},\ldots,a^{\numagents}_{dt})$ for all agents for the current sub-action space $d$.

\subsection{Sequential Action Selection from Joint Logit-Space}
\label{sec:seq_action_selection}

Given the logits of the joint action space $\logits_d \in \mathbb{R}^{\numagents \times |\mathcal{A}_d|}$ for subspace $d$, we iteratively select actions for each agent using common decoding strategies, such as greedy selection or sampling.
To ensure feasibility, joint agent actions are initialized with a set of feasible default actions. For picker locations, the default action is to remain at the current position. Additionally, we introduce a dummy SKU that serves as the default SKU action and can be always selected. 

After each agent received a default action, we iteratively refine the agent actions based on the logits $\logits$. Therefore, we first mask infeasible actions in $\logits$ by setting their values to negative infinity. The masked logits are then converted into a single probability distribution over both agents and actions, rather than creating separate distributions per agent. This approach allows the policy to implicitly learn the ranking $\PermOverL$ by assigning higher logits to agents that should act first. As a result, agents with greater confidence select actions earlier, while less confident agents act later. This step is crucial, as an agent's action can constrain the action space of others, and the order of selection directly affects the picking quantity $y^m$, as detailed in \Cref{sec:mdp}.

Given the probability distribution over the joint action space, a single action (i.e. agent-action combination) $a^m$ is drawn. As a consequence, all actions of the chosen agent $m$ are marked as infeasible. Further, more actions can be masked based on the actions taken so far using a sub-action specific masking function $\xi_d$. 
This way, we avoid that multiple agents select the same shelf- and SKU-combination as required per our MMDP formulation in \Cref{sec:mdp}.
In the next iteration, the logits are computed with the updated action mask and the process repeats until no more actions can be selected (i.e., when all actions are marked infeasible). \Cref{alg:seq-action-selection} formally describes this process.

\definecolor{lightgray}{gray}{0.5}

\begin{algorithm}[t]
\caption{Sequential Action Selection from Joint Logit-Space}
\label{alg:seq-action-selection}
\begin{spacing}{1.0}  
\begin{algorithmic}[1]
\REQUIRE Logits $\logits \in \mathbb{R}^{\numagents \times |\mathcal{A}_d|}$, Binary Action Mask $\mathbf{M} \in \mathbb{R}^{\numagents \times |\mathcal{A}_d|}$, Mask Update Function $\xi_d$, Temperature $\beta$, default action $\bm{a}^\prime$ (e.g. $\bm{a}^\prime \equiv \bm{a}_{t-1}$)
\ENSURE Feasible agent actions $\bm{a} \in \mathbb{N}^\numagents$
\STATE $\bm{a} \gets \bm{a}^\prime$ \hfill \textcolor{lightgray}{\COMMENT{Initialize agent actions}}
\WHILE{not all elements in \( \textbf{M} \) are 1}
    \STATE $\logits^\prime \gets \logits -  \textbf{M} * \infty$ \hfill \textcolor{lightgray}{\COMMENT{Mask infeasible actions}}
    \vspace{1mm}
    \STATE $P_{\agentidx a} \gets \frac{\text{exp}(\logits^\prime_{ma} / \beta)}{\sum_{i\in \SetOfAgents} \sum_{j\in \mathcal{A}_d} \exp(\logits^\prime_{ij} / \beta)} \quad \forall \, \agentidx \in \SetOfAgents, a \in \mathcal{A}_d$ \hfill \textcolor{lightgray}{\COMMENT{Normalize}} \vspace{1mm}
    \STATE $a^\agentidx \sim \text{Categorical}(P)$ \hfill \textcolor{lightgray}{\COMMENT{Sample a single agent's action}}
    \STATE $\bm{a}[\agentidx] = a^\agentidx$ \hfill \textcolor{lightgray}{\COMMENT{Update vector of actions}}
    \STATE \( \textbf{M}[\agentidx, :] = 1 \) \hfill  \textcolor{lightgray}{\COMMENT{Mark all actions of agent \( \agentidx \) infeasible}}
    \STATE $\textbf{M} \gets \xi_d(\textbf{M}, \bm{a})$ \hfill \textcolor{lightgray}{\COMMENT{Update Mask according to subspace specific logic}}
\ENDWHILE
\end{algorithmic}
\end{spacing}
\end{algorithm}

\subsection{Learning Method}

\label{sec:training}
During training, the objective is to adjust the parameters $\theta$ of the policy $p_\theta$ to maximize the reward $R(\bm{a}, x)$ for any given problem instance $x$. Formally, we can cast the optimization problem as follows:
\begin{align}
\label{eq:training_problem_simple_ar}
    \theta^{*}=\underset{\theta}{\text{argmax}} 
    \Big[
    \mathbb{E}_{x \sim P(x)}\big[\mathbb{E}_{\bm{a} \sim p_\theta(\bm{a}|x)}R(\bm{a},x)\big] 
    \Big].
\end{align}

Due to the absence of large datasets containing the optimal solutions $\bm{a}$ for CO problem instances $x$, several Reinforcement Learning techniques have been developed to train neural CO solvers \cite{koolBuyREINFORCESamples2019,kwonPOMOPolicyOptimization2021,kim2022sym}. However, recently, self-supervised approaches have emerged in the realm of neural combinatorial optimization and already achieve state-of-the-art results on some CO problems \cite{pirnay2024selfimprovement,SelfJSP}. 

A major advantage over REINFORCE-based learning is that single actions instead of entire trajectories can serve as training examples. While REINFORCE prohibits a re-encoding of the problem state after each decoding step due to the accumulation of gradient information during training, the use of single actions or sub-trajectories in self-supervised learning allows for stepwise encoding \cite{pirnay2024selfimprovement}. 
While this might impose unnecessary computational cost for static problems like the TSP, it is a strong benefit for a highly dynamic problem like the MSPRP, where after each step the demand, supply and capacity change. Therefore, in this work, we adopt the self-improvement approach described in \cite{pirnay2024selfimprovement}. This method samples $\alpha \gg 1$ candidate solutions for an instance $x$ from the current best-known policy $p_{\theta^\ast}$ and selects the best one, $\bm{a}^\ast \coloneq \text{argmax} \{R(\bm{a}^1, x), \ldots, R(\bm{a}^\alpha, x)\}$, as a training example. Then, cross-entropy loss $\mathcal{L}_{\text{CE}} = - \sum_{t=1}^T \log \, p_\theta (a^\ast_t | s_t)$ is used to train the model on these pseudo-optimal solutions. The refined model is used in the next iteration to generate new candidate solutions, leading to progressively better training examples as training advances.\footnote{A detailed description of the algorithm is given in \Cref{algo:learning}}

In order to apply self-improvement to learn the parameters of MAHAM, we first revise the autoregressive policy of \Cref{eq:autoregressive_encoding_decoding} and extend it with the components introduced by our MAHAM architecture: 

\begin{align}
    p(\bm{a}|x) = \prod_{t=1}^T f_\theta(H|s_t) \cdot \prod_{d=\{\mathcal{V}, \mathcal{P}\}}  g^d_\theta(\logits_d|s_t, H) \cdot \prod_{\agentidx=1}^\numagents \psi(a_{dt}^{m} | \logits_d, \bm{a}_{dt}^{1:m-1}),
\end{align}
where the encoder is factorized over the action sub-spaces $|\mathcal{V}|$ and $|\mathcal{P}|$, which both use the same encoder embeddings, and the decoder produces logits $\logits_d$ only once for all $M$ agents, allowing MAHAM to efficiently model dependencies in multi-agent decision-making. Resulting from this, we train the model with cross entropy loss via gradient descent using the following definition of the gradients:
\begin{align}
    \nabla_\theta \mathcal{L} = - \sum_{t=1}^T \sum_{d=\{\mathcal{V}, \mathcal{P}\}} \sum_{\agentidx=1}^\numagents \nabla_\theta \log \, p_\theta(a_{dt}^{m} | \bm{a}_{dt}^{1:m-1}, s_t)
\end{align}

\section{Experiments}

\label{sec:exp}

We study the effectiveness of MAHAM in solving the min-max MSPRP by comparing it with both traditional OR solvers as well as other multi-agent neural solvers. First, we use the exact solver Gurobi with two different time budgets (10 minutes and one hour) to solve a single instance from the test set. Further, due to the absence of (meta-)heuristics for the min-max variant of the MSPRP, we implement a greedy heuristic as a simple baseline. To compare MAHAM with other learning-based methods, we include HAM \cite{luttmann2024neural}, 2d-Ptr \cite{liu20242d}, Equity Transformer \cite{son2024equity}, and PARCO \cite{berto2024parco} in the experiments.  We describe all baseline solvers in \Cref{appendix:baselines}.

\subsection{Comparison with Baselines}
\newcommand{\best}[1]{\textbf{#1}}
\newcommand{\bstnrl}[1]{#1}

\begin{table*}[t]
\caption{Comparison of MAHAM with baseline solvers.}
\centering
\resizebox{\textwidth}{!}{
\renewcommand\arraystretch{1}
\setlength{\tabcolsep}{3.5mm}
\small{
\begin{tabular}{lccc|ccc|ccc}
\toprule[0.5mm]

\multicolumn{10}{c}{MSPRP10 ($|\mathcal{V}|=10$)}                                                                                                                                                                                 \\ \hline
\multicolumn{1}{c|}{$|\mathcal{P}|$}                              & \multicolumn{3}{c|}{3}                                  & \multicolumn{3}{c|}{6}                                  & \multicolumn{3}{c}{9}                                 \\ \hline
\multicolumn{1}{c|}{Metric}                          & Obj.            & Gap           & \multicolumn{1}{c|}{Time} & Obj.            & Gap       & \multicolumn{1}{c|}{Time} & Obj.            & Gap       & \multicolumn{1}{c}{Time} \\ \hline
\multicolumn{1}{l|}{Gurobi (10m)}                  & \textbf{1.1675}      & 0.0\%        & 28.2s          & \textbf{1.6866}          & 0.0\%              & 381.5s               & \textbf{1.6187}       & 0.0\%        & 249.5s       \\
\multicolumn{1}{l|}{Gurobi (1h)}                  & \textbf{1.1675}      & 0.0\%        & 28.2s          & \best{1.6866}          & 0.0\%              & 381.5s               & \best{1.6187}       & 0.0\%        & 249.5s       \\
\multicolumn{1}{l|}{Greedy}                          & 1.2536           & 7.37\%         & \textbf{0.10s}              & 1.7853               & 5.85\%              & \textbf{0.23s}                  & 1.7311              & 6.94\%          & \textbf{0.24s}        \\\hline
\multicolumn{1}{l|}{HAM}                             & 1.1678            & 0.03\%          & 0.32s              & 1.7089                &   1.32\%          & 0.37s                   & 1.6426                & 1.48\%         &  0.44s \\
\multicolumn{1}{l|}{Equity Trans.}                  & 1.1678         & 0.03\%          & 0,30s          & 1.6903          & 0.22\%              & 0.35s               & 1.6351       & 1.01\%        & 0.43s       \\
\multicolumn{1}{l|}{2d-Ptr}                  & \textbf{1.1675}         & 0.00\%        & 0,32s          & 1.6967          & 0.60\%              & 0.38s               & 1.6371       & 1.14\%        & 0.46s       \\
\multicolumn{1}{l|}{PARCO}                  & \textbf{1.1675}         & 0.00\%         & 0,25s          & 1.6888          & 0.13\%              & 0.31s               & 1.6237       & 0.31\%        & 0.30s       \\
\multicolumn{1}{l|}{MAHAM}                  & \textbf{1.1675}          & 0.0\%         & 0,21s          & 1.6867       &   0.01\%        & 0.25s          &   \best{1.6187}             & 0.0\%        & 0.27s  \\
\midrule[0.4mm]
\multicolumn{10}{c}{MSPRP25 ($|\mathcal{V}|=25$)}       \\ \hline
\multicolumn{1}{c|}{SKUs} & \multicolumn{3}{c|}{12} & \multicolumn{3}{c|}{15} & \multicolumn{3}{c}{18} \\ \hline
\multicolumn{1}{c|}{Metric} & Obj. & Gap & \multicolumn{1}{c|}{Time} & Obj. & Gap & \multicolumn{1}{c|}{Time} & Obj. & Gap & \multicolumn{1}{c}{Time} \\ \hline
\multicolumn{1}{l|}{Gurobi (10m)}           & 1.7608 & 1.42\% & 600s & 1.8402 & 2.87\% & 600s & 1.8929 & 4.22\% & 600s \\
\multicolumn{1}{l|}{Gurobi (1h)}            & 1.7395 & 0.19\%    & 3512s         & 1.7915    & 0.15\%         & 3600s        & 1.8301 & 0.77\% & 3600s \\
\multicolumn{1}{l|}{Greedy}                 & 3.3079      & 90.53\%     & 0.50s         & 2.9636          & 65.68\%          & 0.52s         & 3.4936 & 92.36\% & 0.57s      \\ \hline
\multicolumn{1}{l|}{HAM}                    & 1.7813      & 2.60\%      & 0.89s         & 1.8685          & 4.46\%          & 1.13s          & 1.8954       & 4.36\%        &  1.12s \\
\multicolumn{1}{l|}{Equity Trans.}          & 1.7750      & 2.23\%      & 0.79s         & 1.8328          & 2.46\%          & 1.12s         & 1.8573       & 2.26\%        & 1.11s       \\
\multicolumn{1}{l|}{2d-Ptr}                 & 1.7508      & 0.84\%      & 1.08s         & 1.8332          & 2.48\%          & 1.13s         & 1.8681       & 2.86\%        & 1.15s       \\
\multicolumn{1}{l|}{PARCO}                  & 1.7447      & 0.49\%      & 0.56s         & 1.8014          & 0.70\%          & 0.49s         & 1.8282       & 0.66\%        & 0.51s       \\
\multicolumn{1}{l|}{MAHAM}                  & \textbf{1.7362} & 0.00\% & \textbf{0.45s} & \textbf{1.7888} & 0.00\% & \textbf{0.46s} & \textbf{1.8162} & 0.00\% & \textbf{0.47s} \\
\midrule[0.5mm]
\multicolumn{10}{c}{MSPRP40 ($|\mathcal{V}|=40$)}  \\ \hline  
\multicolumn{1}{c|}{$|\mathcal{P}|$}        & \multicolumn{3}{c|}{15} & \multicolumn{3}{c|}{20} & \multicolumn{3}{c}{30} \\ \hline
\multicolumn{1}{c|}{Metric} & Obj.          & Gap & \multicolumn{1}{c|}{Time} & Obj. & Gap & \multicolumn{1}{c|}{Time} & Obj. & Gap & \multicolumn{1}{c}{Time} \\ \hline
\multicolumn{1}{l|}{Gurobi (10m)}           & 1.9163            & 17.26\%       & 600s          & 2.1907            & 12.00\%            & 600s             & 2.3398            & 32.99\%       & 600s \\
\multicolumn{1}{l|}{Gurobi (1h)}            & 1.7552            & 7.40\%        & 3600s          & 2.0201           & 3.28\%            & 3600s                & 1.8699         & 6.28\%        & 3600s \\
\multicolumn{1}{l|}{Greedy}                 & 4.1010            & 150.93\%      & 0.63s          & 5.1602            & 163.81\%          & 1.02s               & 4.0420         & 129.74\%      & 1.12s \\ \hline
\multicolumn{1}{l|}{HAM}                    & 1.7256            & 5.59\%        & 1.31s           & 2.1334           & 9.07\%           & 1.71s                & 1.9211       & 9.19\%          & 3.66s \\
\multicolumn{1}{l|}{Equity Trans.}          & 1.6985            & 3.93\%        & 1.60s          & 2.0373          & 4.16\%             & 2.16s               & 1.8355       & 4.33\%           & 3.77s     \\
\multicolumn{1}{l|}{2d-Ptr}                 & 1.6857            & 3.15\%        & 1.65s          & 2.0245          & 3.50\%             & 2.16s               & 1.8232       & 3.63\%           & 2.90s     \\
\multicolumn{1}{l|}{PARCO}                  & 1.6452            & 0.67\%        & 0.72s          & 1.9760          & 1.02\%             & 0.79s               & 1.7896       & 1.72\%           & 1.16s     \\
\multicolumn{1}{l|}{MAHAM}                  & \textbf{1.6343} & 0.00\% & \textbf{0.54s} & \textbf{1.9560} & 0.00\% & \textbf{0.66s} & \textbf{1.7594} & 0.00\%   & \textbf{0.92s}     \\
\bottomrule[0.5mm]
\end{tabular}
}}
\label{tab:main}
\end{table*}
We present the main empirical results, comparing MAHAM against all baselines mentioned above, in \Cref{tab:main}, reporting the average objective function values (Obj.), gaps to the best-known solutions, and inference times for solving a single instance from the test set of the respective instance type. For training and evaluating MAHAM, we use the same instance types and instance generation method as described in \cite{luttmann2024neural}. Specifically, we use three different warehouse layouts, with 10, 25, and 40 shelves and vary the number of SKUs per layout type. We describe the generation of instances in detail in \Cref{appendix:instance}. For neural baselines, we evaluate the performance using 1280 sampled solutions and reporting the objective value of the best one. 

MAHAM consistently outperforms other neural baselines in terms of solution quality and speed, with margins growing with the size of the problem instance. Also, MAHAM is on-par with the Gurobi solver on small instances and even outperforms it on larger instances, where no optimal solutions were found in the given time bounds.

\subsection{Large Scale Generalization}
\begin{table*}[t]
\centering
\caption{Large-scale generalization for unseen numbers of locations and SKUs}
\label{table:large_scale_generalization}
\resizebox{\textwidth}{!}{
\renewcommand\arraystretch{1}
\setlength{\tabcolsep}{3.5mm}
\small{
\begin{tabular}{l|ccc|ccc|ccc}
\toprule[0.5mm]
\multicolumn{10}{c}{MSPRP50 ($|\mathcal{V}|=50$)} \\ 
\hline
\multicolumn{1}{c|}{$|\mathcal{P}|$} & \multicolumn{3}{c|}{100} & \multicolumn{3}{c|}{250} & \multicolumn{3}{c}{500} \\ 
\hline
Method          & Obj.               & Gap     & Time    & Obj.             & Gap    & Time         & Obj.               & Gap    & Time         \\  \hline
Greedy          & 4.9638             & 122\%   & 1.75s   & 5.7146           & 85\%   & 6.22s        & 6.0632             & 51\%   & 28.77s       \\ 
2d-Ptr          & 3.9799             & 78\%    & 5.56s   & 5.9040           & 91\%   & 19.62s       & 6.6315             & 65\%   & 61.85s       \\ 
PARCO           & 3.9412             & 76\%    & 3.98s   & 5.1629           & 67\%   & 10.70s       & 5.2790             & 32\%   & 27.20s       \\ 
MAHAM w/o PS    & 2.3865             & 7\%     & 3.68s   & 3.1643           & 2\%    & 8.32s        & 4.1617             & 4\%    & 17.55s       \\ 
MAHAM           & \textbf{2.2352}    & 0\%     & 3.55s   & \textbf{3.0916}  & 0\%    & 7.79s        & \textbf{4.0128}    & 0\%    & 15.10s       \\ 
\bottomrule[0.4mm]
\end{tabular}
}
}
\end{table*}

We further evaluate the generalization performance of MAHAM on large-scale instances of the MSPRP that were not seen during training. The ability to generalize to larger instances is crucial for any NCO algorithm to make it applicable to dynamic real-world scenarios. We evaluate MAHAM against a purely autoregressive approach (2d-Ptr), PARCO as an alternative parallel decoding model, and the Greedy heuristic. Gurobi is not included in the evaluation as it can not find solutions to any of the large instances with a time budget of one hour per instance. The results are shown in \Cref{table:large_scale_generalization}, where MAHAM consistently outperforms other methods while also being significantly faster. Most notable is the large performance gap compared to PARCO, which achieves competitive results in in-distribution testing, but seems to generalize worse to larger instances.

\subsection{Ablation Studies}

\paragraph{Picker Ranking and Coordination:}
A key aspect of MAHAM is the Sequential Action Selection. The quality of the solution generated by the autoregressive policy defined in \Cref{eq:maham} and the action selection function $\psi$ defined by \Cref{alg:seq-action-selection} strongly depends on the order $\PermOverL$ in which pickers perform actions. In order to validate that the model is able to learn good agent rankings, i.e., by assigning higher logits to those agents that should have priority, we compare the (\textit{learned}) agent priority of \Cref{eq:maham} with a model that iterates over the set of agents in the order of their index $m$ (\textit{index}) as well as a model that determines the order $\Omega$ randomly (\textit{random}). 

Moreover, MAHAM utilizes a separate agent encoder that enables effective agent coordination when computing the joint action logits in the decoders. The idea of utilizing an agent encoder for multi-agent problems itself is not novel, but has already been applied in the Equity Transformer \cite{son2024equity}, the 2d-Ptr \cite{liu20242d}, and PARCO \cite{berto2024parco}. However, in this work we fuse the agent encoder with a novel rank-dependent positional encoding followed by a multi-head self-attention layer. This enables effective communication between the agents based on their current utilization and ultimately enables the model to come up with optimal rankings. 

\begin{figure}[h]
    \centering
    \scalebox{0.95}{ 
        \begin{minipage}{\textwidth}
            \centering
            \begin{subfigure}[b]{0.435\textwidth}
                \centering
                \includegraphics[width=\textwidth]{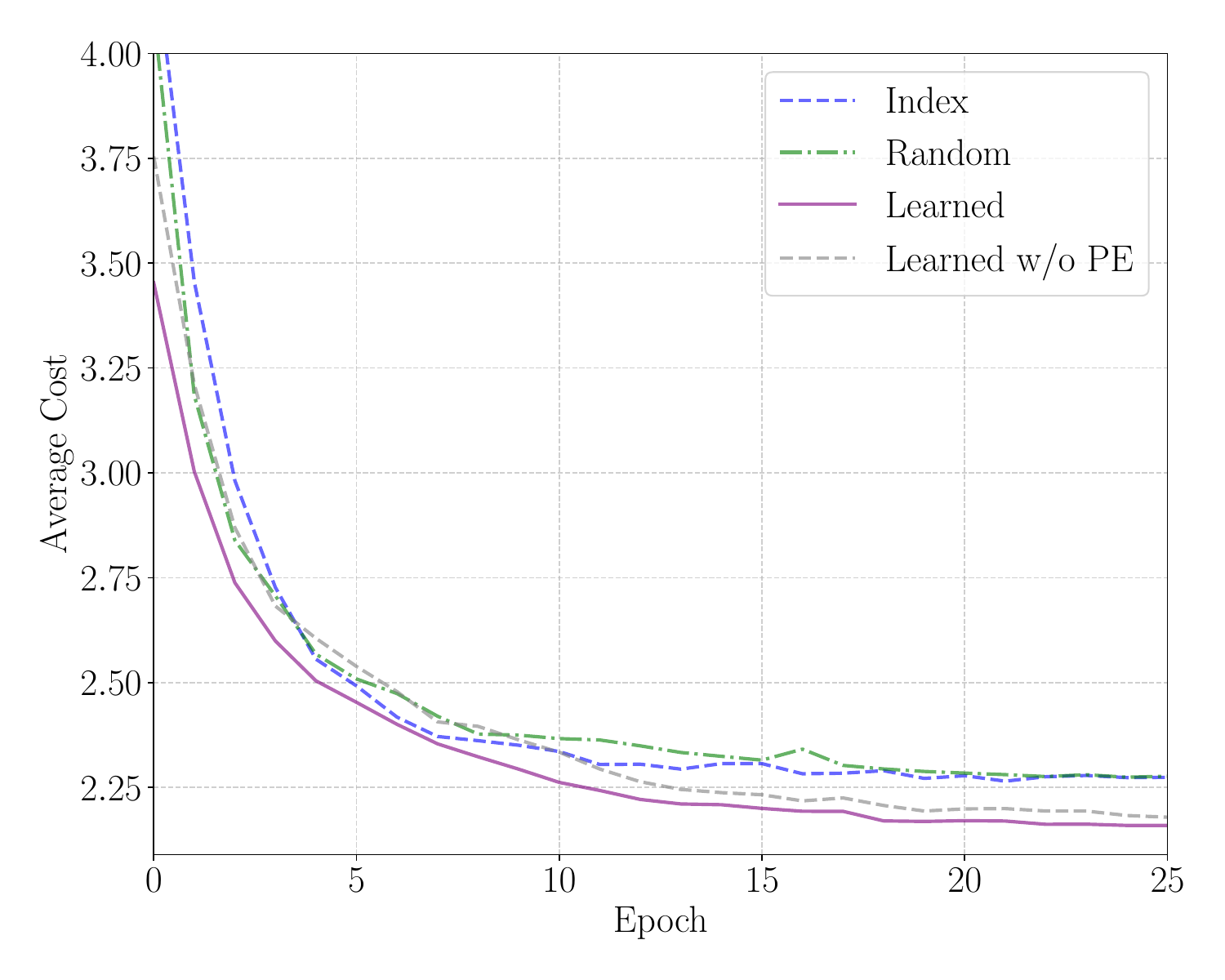}
                \vspace{-6mm}
                \caption{}
                \label{fig:ranking}
            \end{subfigure}
            \hfill
            \begin{subfigure}[b]{0.5\textwidth}
                \centering
                \includegraphics[width=\textwidth]{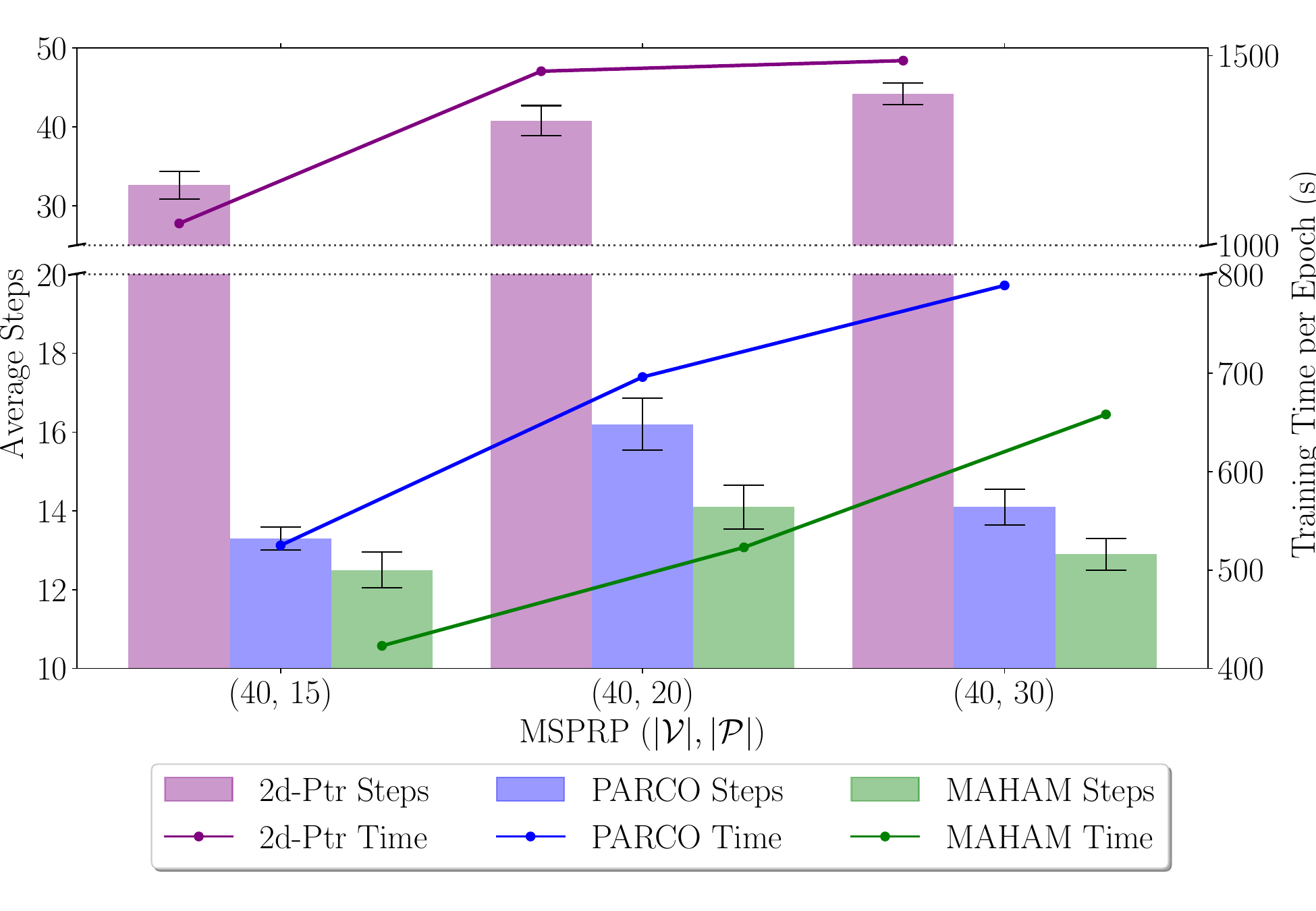}
                \vspace{-6mm}
                \caption{}
                \label{fig:runtime}
            \end{subfigure}
        \end{minipage}
    }
    \caption{\footnotesize Solution quality of MAHAM on MSPRP instances for different ranking strategies (left) and MAHAM efficiency in comparison to the 2d-Ptr and PARCO (right)}
    \label{fig:both}
\end{figure}

\Cref{fig:ranking} summarizes the results of an ablation study testing the effectiveness of the proposed components in our MAHAM architecture. The full model with learned rankings and rank-dependent positional encodings (PE) performs significantly better than the models relying on an index-based or random order, and also achieves better solutions than MAHAM without the positional encoding.

\paragraph{Encoder Parameter Sharing:} MAHAM introduces an efficient way to incorporate message-passing over different types of nodes in a heterogeneous graph. Through the parameter sharing (PS) approach described in \Cref{sec:encoder}, the MAHAM encoder saves roughly 20\% in size, allowing it to process larger instances faster. In addition, parameter sharing acts as regularization, improving the generalization of the trained model. We compare MAHAM with and without parameter sharing in the cross-attention layer of the encoder on out-of-distribution instances in \cref{table:large_scale_generalization}. Parameter sharing consistently results in better solutions in less time.  

\vspace{-5mm}
%
%
\subsection{Runtime Comparison}
We study the efficiency of MAHAM by comparing it to the 2d-Ptr -- acting as a purely autoregressive neural baseline -- and PARCO, another parallel solution construction approach. The results are shown in \Cref{fig:runtime}. While the 2d-Ptr requires much more decoding steps to construct a solution, resulting in longer training times, MAHAM also needs less construction steps and is quicker to train than PARCO. This can be attributed to our Sequential Action Selection approach, which effectively avoids conflicts through adaptive masking. In PARCO on the other hand, agents may select the same shelf/SKU in the same decoding step, resulting in a conflict and consequently in one or more agents doing nothing in the respective stage.


\section{Conclusion and Future Work}
In this work, we introduced the first neural solver for the min-max Mixed-Shelves Picker Routing Problem. The core of our approach is the integration of a hierarchical and parallel decoding mechanism capable of efficiently constructing solutions over complex, multi-dimensional action spaces, such as those found in min-max MSPRP. While previous methods relied on sequential solution construction or parallel decision-making prone to conflicts, our approach achieves efficient and effective agent coordination, enabled by a novel Sequential Action Selection algorithm.

Our extensive experimental results, including traditional as well as neural solvers, demonstrate the superiority of MAHAM in both solution quality and inference speed, particularly for large-scale problem instances. These findings highlight the capabilities of neural solvers and prove them as a strong alternative to hand-crafted heuristics.

Future research directions include extending this approach to more dynamic warehouse environments with real-time demand fluctuations and exploring hybrid methods that integrate learning-based techniques with optimization heuristics for further performance improvements. Additionally, our framework could be adapted to other multi-agent combinatorial optimization problems beyond warehouse logistics, such as fleet routing and robotic task allocation.

\bibliographystyle{splncs04}
\bibliography{bib.bib}

\clearpage

\newpage
\appendix
\onecolumn
\renewcommand{\paragraph}[1]{\bigskip\noindent\textbf{#1}\quad} 

\section{Formal Definition of the MSPRP}
\label{appendix:notation}

The following mathematical model describes the min-max MSPRP cover in this work and table \ref{tab:notation} summarizes the notation used to define the model.

\begin{align}
  \label{eq:obj_fn}
  \textbf{min} \qquad \qquad \qquad \quad Z &= \underset{b \in \NumTours}{\text{max}} \sum_{(i,j) \in \mathcal{E}} D_{ij} \cdot x_{ijb} \\
    \label{eq:flow_cnstr}
    \textbf{s.t.} \qquad \quad \; \; \sum_{(i,j) \in \mathcal{E}} x_{ijb} &=  \sum_{(j,i) \in \mathcal{E}} x_{jib} &&\forall \, i \in \SetOfAllNodes, b \in \NumTours \\
    \label{eq:no_mult_visits}
    \sum_{(i,j) \in \mathcal{E}} x_{ijb} &\leq 1 &&\forall \, i \in \SetOfAllNodes, \, b \in \NumTours \\
    \label{eq:take_only_if_visit}
    BigM \cdot \sum_{i \in \SetOfAllNodes} x_{ijb} &\geq y_{jb} && \forall \, j \in \SetOfStorageLocations , \,b \in \NumTours \\
    \label{eq:include_depot}
    \sum_{h \in \SetOfPackingStations} \sum_{j \in \SetOfStorageLocations} x_{hjb} &= 1 && \forall \, b \in \NumTours \\
     \label{eq:subtour}
    \sum_{i\in S} \sum_{j \in S} x_{ijb} &\leq |S| -1 && \forall \, b \in \NumTours, \,  S \subset \SetOfStorageLocations, |S| \geq 2 \\
    \label{eq:capacity}
     \sum_{i \in \SetOfStorageLocations} y_{ib} &\leq \kappa && \forall b \in \NumTours\\
    \label{eq:meet_demand}
    \sum_{i \in \SetOfLocationsIncludePickItem{p}} \sum_{b \in \NumTours} y_{ib} &= d_p && \forall \, p \in \SetOfSKUs \\
    \label{eq:no_exceed_sup}
    \sum_{b \in \NumTours} y_{ib} &\leq n_{i} && \forall \, i \in \SetOfStorageLocations \\
    \label{eq:def_x}
     x_{ijb} &\in \{0,1\} && \forall \, (i,j) \in \mathcal{E},\, b \in \NumTours \\
    \label{eq:def_y}
     y_{ib} &\geq 0 && \forall \,  i \in \SetOfStorageLocations, \, b \in \NumTours
\end{align}

The objective function (\ref{eq:obj_fn}) aims to minimize the maximum distance traveled by any picker. Constraints (\ref{eq:flow_cnstr}) ensure that every storage location visited during a picker’s tour is also exited. Constraints (\ref{eq:no_mult_visits}) prevent storage locations from being visited multiple times within a single tour, though multiple visits are allowed across tours if the picker's capacity is insufficient to fulfill the demand in one trip. However, revisiting the same storage location within a single tour is inefficient and therefore disallowed.
Using the Big-M formulation in (\ref{eq:take_only_if_visit}), we ensure that items can only be picked from storage locations included in the respective tour. Since no more than $\kappa$ items can be picked in one tour, setting $BigM=\kappa$ is sufficient.
To guarantee that each tour begins and ends at a packing station, constraints (\ref{eq:include_depot}) enforce that a packing station is exited exactly once per tour. Combined with the network flow constraints (\ref{eq:flow_cnstr}), this ensures that every tour returns to the packing station it initially departed from. Additionally, subtour elimination constraints (\ref{eq:subtour}) ensure that all visited storage locations are connected within a tour.
Constraints (\ref{eq:capacity}) prevent the picker's capacity from being exceeded, while constraints (\ref{eq:meet_demand}) ensure that all customer orders are fulfilled. To avoid exceeding the available stock of items in any storage location during order picking, constraints (\ref{eq:no_exceed_sup}) are enforced. Finally, constraints (\ref{eq:def_x}) and (\ref{eq:def_y}) define the domains of the decision variables $x$ and $y$.
\begin{table}
    \centering
    \caption{Notation used in the MIP-Model}
    \label{tab:notation}
    \renewcommand{\arraystretch}{1.1}
    \begin{tabularx}{\columnwidth}{>{\hsize=0.1\columnwidth}X @{\hskip 3pt}|@{\hskip 5pt} >{\hsize=0.87\columnwidth}X}
        \hline
        Symbol & \multicolumn{1}{c}{Description} \\ \hline
        $\SetOfSKUs$ & Set of SKUs for picking\\
        $\SetOfAllNodes$ & Set of storage locations and packing stations $(\SetOfAllNodes=\SetOfStorageLocations \cup \SetOfPackingStations)$\\
        $\mathcal{E}$ & Set of edges $\{(i,j) \, | \, i,j \in \SetOfAllNodes, \, i \neq j  \} $\\
        $\SetOfLocationsIncludePickItem{p}$ & Set of storage locations including picking item $p \in \SetOfSKUs$ \\
        $\NumTours$ & Set of required tours $\{1,2,...,|\NumTours|\}$\\
        $D_{ij}$ & Distance between two nodes $(i,j) \in \mathcal{E}$\\
	$\kappa$ & Maximum picking capacity per tour\\
	$d_{p}$ & Total demand for item $p \in \SetOfSKUs$\\
	$n_{i}$ & Available supply at storage location $i \in \SetOfStorageLocations$\\
        $x_{ijt}$ &  Binaray variable, indicating whether node $j \in \SetOfAllNodes$ has been visited after node $i \in \SetOfAllNodes$ in tour $b \in \NumTours$ \\
        $y_{it}$ & Units picked up at location $i \in \SetOfStorageLocations$ in tour $b \in \NumTours$\\\hline
    \end{tabularx}

\end{table}

\section{Baselines}
\label{appendix:baselines}

\paragraph{Gurobi \cite{gurobi}.}
We implement the mathematical model described in \Cref{appendix:notation} in the exact solver Gurobi \cite{gurobi} and provide a time budget of 600 and 3600 seconds per test instance. We run the Gurobi solver with activated multi-threading on a machine equipped with two Intel Xeon E5-2690 v4 processors, totaling 28 physical cores and 56 logical threads.

\paragraph{Greedy.}
Due to the absence of heuristics developed for the min-max MSPRP, we develop a greedy heuristic as a simple baseline. The heuristic constructs solutions sequentially by assigning each agent logits for selecting a shelf, weighted inversely by its distance from the agent's current position. Similarly, SKUs are chosen with logits proportional to the number of units an agent could potentially pick. Given the logits, the same sequential action selection as described in \Cref{alg:seq-action-selection} is used to generate actions for all agent. Being a stochastic heuristic, we use it to generate 100 different solutions for each test instance and select the best one.

\paragraph{Hierarchical Attention Model \cite{luttmann2024neural}.}
The Hierarchical Attention Model (HAM) introduces the idea of a hierarchical decoder to generate actions over the decomposed action space of the MDP formulation of the MSPRP. Although HAM was developed to solve the min-sum MSPRP, creating $\NumTours$ one after another, it can be used to solve the min-max MSPRP as well thanks to our assumption, that there are exactly as many pickers as there are tours. In this work, HAM is trained like all other models on the min-max-based reward defined in \Cref{sec:mdp} using the learning method outlined in \Cref{sec:training}.

\paragraph{2d-Ptr \cite{liu20242d}.}
The 2D Array Pointer network (2d-Ptr) addresses the heterogeneous capacitated vehicle routing problem (HCVRP) by using a dual-encoder setup to map vehicles and customer nodes effectively. This approach facilitates dynamic, real-time decision-making for route optimization. Its decoder employs a 2D array pointer for action selection, prioritizing actions over vehicles. The 2d-Ptr can be adapted to solve the min-max MSPRP by using the 2D pointer hierarchically to select shelves and SKUs and by using pickers instead of vehicles.

\paragraph{Equity Transformer (ET) \cite{son2024equity}.}
The Equity-Transformer (ET) approach  \cite{son2024equity} addresses min-max routing problems by employing a sequential planning approach with sequence generators like the Transformer. It focuses on equitable workload distribution among multiple agents, applying this strategy to challenges like the min-max multi-agent traveling salesman and pickup and delivery problems. In our experiments, we modify the agent context in the decoder to the MSPRP setting

\paragraph{PARCO \cite{berto2024parco}.}
PARCO is a recent NCO framework for solving multi-agent CO problems. It uses a multi-pointer mechanism paired with a conflict handler to generate solutions for multiple agents in parallel. It is a versatile framework, which has been applied to different routing and scheduling problems.

\section{Model And Training Configuration}

In the following, we detail the model and training parameters as well as the parameters for generating the training / test data. Besides that, to ensure proper reproducibility we provide all training details in our publicly available GitHub repository as configuration files.

\subsection{Instance Generation}
\label{appendix:instance}
For training and evaluating MAHAM and the baselines described above, we use the same instance generation scheme described in \cite{luttmann2024neural}, who generate instances for three warehouse types that differ in the number of available shelves. They generate instances with 10, 25 and 40 shelves referred to as MSPRP10, MSPRP25 and MSPRP40, respectively. While the number of shelves is fixed, the number of demanded SKUs is altered for each warehouse type.

We randomly select the $|\SetOfStorageLocations|$ storage locations from all $|\SetOfSKUs| \times |\mathcal{V}^{\mathrm{R}}|$ possible SKU-shelf combinations and sample the supply from a discrete uniform distribution with mean $\Bar{n}_i$. Likewise, the demand for each SKU is sampled from a discrete uniform distribution with mean $\Bar{d}_p$. Lastly, we clip the demand of an SKU by the warehouse's total supply for it in order to ensure the feasibility of all generated instances. 
Table \ref{tab:instances} summarizes the parameters of the different instances.

\begin{table}[h]
\renewcommand{\arraystretch}{1.1} 
\centering
\caption{Parameter values for instance generation}
\label{tab:instances}
\setlength{\tabcolsep}{10pt} 
\resizebox{\textwidth}{!}{
\begin{tabular}{p{1.5cm}|ccc|ccc|ccc|ccc}
\toprule
                             & \multicolumn{3}{c|}{MSPRP10}    & \multicolumn{3}{c|}{MSPRP25}   & \multicolumn{3}{c|}{MSPRP40} & \multicolumn{3}{c}{MSPRP50} \\ 
\midrule
$|\mathcal{V}^\mathrm{R}|$   & 10  & 10  & 10  & 25  & 25  & 25  & 40  & 40  & 40 & 50  & 50  & 50  \\
$|\mathcal{V}^\mathrm{S}|$   & 20  & 20  & 20  & 50  & 50  & 50  & 100  & 100  & 100 & 200  & 500  & 1000  \\
$|\SetOfSKUs|$               & 3   & 6   & 9   & 12  & 15  & 18  & 15  & 20  & 30 & 100  & 250  & 500  \\ 
$\kappa$                     & 6   & 9   & 9   & 12  & 12  & 15  & 12  & 15  & 15 & 15  & 15  & 15  \\ 
\bottomrule
\end{tabular}
}
\end{table}

\subsection{Network Hyperparameters}
\label{appendix:network}
To ensure valid and meaningful experiments, the hyperparameters are identical for all models. The size of the embeddings is set to 256 and the number of heads for multi-head attention mechanisms is set to 8. All models use $L=4$ encoder layers, GELU activation functions \cite{hendrycks2016gaussian}, and Layer Normalization \cite{ba2016layernormalization}. To map the different entities of the MSPRP into embedding space, all models use the same features outlined in \Cref{tab:features}.

\begin{table}[H]
\centering
\setlength\tabcolsep{5pt}
\caption{Features to describe the different entities in the min-max MSPRP}
\label{tab:features}
\renewcommand{\arraystretch}{1.1}
\begin{tabular}{p{0.4cm}|l}
\hline
\multicolumn{1}{c|}{$\phi$} & \multicolumn{1}{c}{Description} \\
\hline
\parbox[t]{2mm}{\multirow{3}{*}{\rotatebox[origin=c]{90}{Station}}} & Cartesian Coordinates of the packing station \\
& Amount of items to be commissioned at the station \\  
& Number of agents belonging to the station \\
\hline
\parbox[t]{2mm}{\multirow{3}{*}{\rotatebox[origin=c]{90}{Shelf}}} & Cartesian Coordinates of the shelf \\
& Number of different SKUs stored in shelf $i$: $n_i = |\{p \in \mathcal{P} \, | \, E_{ip} > 0 \}|$ \\
& The average supply for SKUs stored in shelf $i$: $\bar{e}_i = (\sum_{p \in \mathcal{P}} E_{ip}) / n_i  $ \\
\hline
\parbox[t]{2mm}{\multirow{3}{*}{\rotatebox[origin=c]{90}{SKU}}} 
& Demand $d_p$ of SKU \\
& Number of shelves the SKU $p$ is available in: $n_p = |\{p \in \mathcal{P} \, | \, E_{ip} > 0 \}|$ \\
& The average storage quantity of SKU $p$: $\bar{e}_p = (\sum_{i \in \mathcal{V}} E_{ip}) / n_p  $ \\
\hline
\parbox[t]{2mm}{\multirow{4}{*}{\rotatebox[origin=c]{90}{Agent}}} 
& Remaining capacity of picker $\kappa^m_t$ \\
& Length of the picker's current tour $dist(\tau^m_{1:t})$ \\
& The total remaining demand of all SKUs $\sum_{p \in \mathcal{P}} d_p$ \\
& The embedding of the agents current location $\bm{h}_\mathcal{V}^m = H_\mathcal{V}[v^m]$ \\
\hline
\end{tabular}
\end{table}

\subsection{Training Hyperparameters}
\label{appendix:training}
We train all models using the self-improvement method described by \cite{pirnay2024selfimprovement}. To ensure consistency, we use identical hyperparameters and training environments for all neural baselines described in \Cref{appendix:baselines}. All models are trained on a single NVIDIA A100 GPU with 40GB of VRAM. Training spans 50 epochs, with each epoch generating $N=5,000$ independent instances. For each instance, $\alpha=100$ candidate solutions are sampled from the reference-policy $\pi_{\text{best}}$, and the best solution is added to the training dataset.
After all instances are solved by the reference policy $\pi_{\text{best}}$, we draw training samples in mini-batches of $B=2.000$ and determine the cross-entropy loss for the pseudo-optimal actions with respect to the target-policy $\pi$. Adam optimizer with a learning rate of 0.0001 is used to update the parameters of the target-policy, and the trainer class from the RL4CO \cite{berto2023rl4co} library is used to guide the learning process.  

The validation dataset consists of 10,000 independently generated instances per epoch. If the target policy outperforms the reference policy on the validation set, the reference policy is updated, and the training dataset is reset. \Cref{algo:learning} provides a detailed breakdown of these steps.
\definecolor{lightgray}{gray}{0.5}

\begin{algorithm}[h]
   \caption{Self-improvement training for neural CO}
   \label{algo:learning}
   \begin{algorithmic}[1]
      \REQUIRE $\mathcal{X}$: distribution over problem instances; $f_*$: objective function
      \REQUIRE $N$: number of instances to sample in each epoch
      \REQUIRE $\alpha$: number of sequences to sample for each instance
      \REQUIRE $\textsc{Validation} \sim \mathcal{X}$: validation dataset
      \STATE Randomly initialize policy $\pi_\theta$
      \STATE $\pi_{\text{best}} \gets \pi_\theta$
      \STATE $\textsc{Dataset} \gets \emptyset$
      \FOR{epoch}
          \STATE Sample set of $n$ problem instances $\textsc{Instances} \sim \mathcal{X}$
          \FOR{each $x \in \textsc{Instances}$}
              
              \STATE \textcolor{lightgray}{// Sample set of $m$ feasible solutions}
              \STATE $A := \{\bm a_{1:T}^{(1)}, \dots, \bm a_{1:T}^{(m)}\} \sim \pi_{\text{best}}$
              \STATE \textcolor{lightgray}{// Add best solution to training dataset}
              \STATE $\textsc{Dataset} \gets \textsc{Dataset} \cup \{(x, \arg\max_{\bm a_{1:T} \in A} f_x(\bm a_{1:T}))\}$ 
          \ENDFOR
          \FOR{batch}
              \STATE \textcolor{lightgray}{// Sample $B$ instances and partial solutions from \textsc{Dataset}}
              \STATE $\{(x_j, \bm{a}^{(j)}_{1:d_j})\}_{j=1}^B \sim \textsc{Dataset}, \qquad \{d_j\}_{j=1}^B \sim \mathcal{U}(1, T-1)$ 
              \STATE \textcolor{lightgray}{// Minimize batch-wise cross entropy loss}
              \STATE $\mathcal{L}_\theta = - \frac{1}{B}\sum_{j=1}^B \log \pi_\theta \left( a^{(j)}_{d_{j+1}} | \bm a^{(j)}_{1:d_j} \right)$
          \ENDFOR
          \IF{greedy performance of $\pi_\theta$ on \textsc{Validation} better than $\pi_{\text{best}}$}
              \STATE \textcolor{lightgray}{// update best policy}
              \STATE $\pi_{\text{best}} \gets \pi_\theta$ 
              \STATE \textcolor{lightgray}{// Empty Training Dataset}
              \STATE $\textsc{Dataset} \gets \emptyset$ 
          \ENDIF
      \ENDFOR
   \end{algorithmic}
\end{algorithm}

\end{document}